\documentstyle[preprint,aps]{revtex}

\newcommand{\bb}{\begin{equation}}
\newcommand{\ee}{\end{equation}}
\newcommand{\bqn}{\begin{eqnarray}}
\newcommand{\eqn}{\end{eqnarray}}
\newcommand{\pp}{\partial }
\newcommand{\m}{\frac{1}{2}}
\newcommand{\w}{\mbox{\tiny $\wedge$}}
\newcommand{\eij}{\epsilon^{ij}}
\newcommand{\K}{\frac{k}{2\pi}}

\newcommand{\Ann}[1]{{\sl Ann.~Phys.}~{\bf #1}}

\newcommand{\PLB}[1]{{\sl Phys.~Lett.}~{\bf B#1}}

\newcommand{\PRL}[1]{{\sl Phys.~Rev.~Lett.}~{\bf #1}}

\newcommand{\IJMPA}[1]{{\sl Int.~J.~Mod.~Phys.}~{\bf A#1}}

\newcommand{\PRD}[1]{{\sl Phys.~Rev.}~{\bf D#1}}

\begin{document}

\title{Global charges in Chern-Simons theory and the 2+1 black hole}

\author{M\'aximo Ba\~nados}

\address{Centro de Estudios Cient\'{\i}ficos de Santiago, Casilla
16443, Santiago 9, Chile \\ and \\ The Blackett Laboratory, Imperial College,
Prince Consort Road, London SW7 2BZ, England. }
\maketitle

\begin{abstract}
We use the Regge-Teitelboim method to treat surface integrals in
gauge theories to find global charges in Chern-Simons theory.
We derive the affine and Virasoro generators as global charges
associated with symmetries of the boundary.  The role of boundary
conditions is clarified. We prove that for
diffeomorphisms that do not preserve the boundary there is a
classical contribution to the central charge in the Virasoro
algebra. The example of anti-de Sitter 2+1 gravity is considered
in detail.
\end{abstract}

\section{Introduction}

Global charges in Chern-Simons theory can be understood by
several independent methods. Perhaps the most popular one is
found in the observation that the Chern-Simons action, defined on
a manifold with boundaries, induces a Wess-Zumino-Witten theory
at the boundary \cite{CS-WZW}. Then, at least for simple
topologies, the symplectic structure associated with the degrees of
freedom at the boundary is the affine extension of
the Lie algebra considered \cite{Witten84}, and their
corresponding Virasoro generators can be constructed by the
Sugawara construction. From the point of view of the classical
theory, which will be our main interest here, this gives an
infinite set of global charges that satisfy a well defined (Dirac
bracket) algebra. [The Virasoro generators are associated to the
diffeomorphism invariance of the theory. They are not
independent from the affine generators because, in Chern-Simons
theory, the local gauge group contains the diffeomorphism group.]

A different approach leading to the same results was followed in
Ref. \cite{Balachandran}.  In that reference, the authors studied the
differentiability properties of the generators of gauge
transformations.  Imposing strong boundary conditions they ensure
their differentiability and then, they find a set of first class
quantities (which are not zero on-shell) satisfying an affine
extension of the Lie algebra. The same procedure is then applied to
gauge transformations that generate diffeomorphisms showing that an
analogous set of first class quantities exist such that they
satisfy the Virasoro algebra.  Both set of first class functions were
shown to be related by the Sugawara construction.  More
recently, the asymptotic group of anti-de Sitter 2+1
gravity using the Chern-Simons formulation was studied in
\cite{Ezawa}. In that work, however, the advantages of the
Chern-Simons formulation are not fully explored because the boundary
conditions are read off from the metric formulation.

In this paper we apply the Regge-Teitelboim
\cite{Regge-Teitelboim} method to treat boundary terms and boundary
conditions in gauge field theory. This method, although closely
related to the one used in \cite{Balachandran}, improves it in
several ways. (1) It deals only with the generators of gauge
transformations and their associated charges, it is not necessary
to introduce other first class quantities. (2) It provides a
natural interpretation for global charges as the
generators of the residual gauge group (in the Dirac
bracket) after the gauge is fixed. [This point was not properly
examined in\cite{Balachandran}.] (3) The boundary conditions are
somehow dictated by the theory and not imposed from the
outside. This allows us to relax the boundary conditions
such that a {\em classical} central charge in the Virasoro
algebra will appear. [This boundary conditions allows for
diffeomorphisms that have a non-zero component normal to the
boundary.]

In section \ref{Global-charges} we study the differentiability
of the generators of gauge transformations and define the charges
that regularize them for a generic choice of boundary conditions.
Then, we consider the two particular cases leading to the affine and
Virasoro algebras.

The stratagy to define global charges in gauge theories is standard
\cite{Regge-Teitelboim}. Given a set of first class constraints $G_a
\approx 0$ satisfying
\bb
\{ G_a, G_b \} = f^c_{\; ab} G_c ,
\label{fc}
\ee
one considers the smeared generator (supplemented by a charge $Q$ that
makes it differentiable)
\bb
G(\eta) = \int \eta^a G_a + Q(\eta) \approx Q(\eta) .
\label{sg}
\ee
Since $G(\eta) $ is differentiable, one can compute its Poisson
bracket with $G(\lambda)$. The question is whether or not the
smeared generators satisfy a closed algebra.  The answer to this
question is intimately related to the issue of boundary
conditions. Indeed, the value of the charge $Q$
crucially depends on them.  Normally, one first chooses a set of boundary
conditions --with precise fall-off conditions for the fields--
and then finds the most general set of gauge transformations
that leave them invariant.  This set of
parameters defines what we call global symmetries. The
smeared generator (\ref{sg}) is then supplemented with the extra
condition that the parameter of the transformation ($\eta$), at
the boundary, must be a global symmetry. It turns out that
global symmetries are not generated by the constraints but by
the complete generator (\ref{sg}) which is not weakly zero. The
conserved charge associated to this symmetry is the charge $Q$,
being also the generator of the global symmetry after the gauge
is fixed.  Global symmetries form, in general, a subgroup of
the complete gauge group.  For example, the global group of
asymptotically flat 3+1 gravity is the Poincare group, which, of
course, do not contain the whole of the diffeomorphism invariance.

Another  problem related to the issue of boundary conditions  is the
gauge fixing procedure in gauge in field theory. Since, in general, the
constraint are
differential functions of the canonical variables, it is not possible
to completely fix the gauge (determine the Lagrange multipliers) by a
canonical gauge condition. One is left, instead, with a differential
equation for the Lagrange multipliers which contains ``integrations
functions" at the boundary. These functions represent the freedom to
make ``improper'' or global gauge transformations, even after the
gauge is fixed.  In section \ref{Gauge-fixing} we describe how to fix
the gauge in the simple case in which the (spatial) manifold has the
topology of a disc.  We shall see that the connection
is fixed up to an arbitrary function at the boundary.  Also, there is a
non-trivial residual symmetry at the boundary.

Once the boundary conditions are chosen and the group of global
symmetries is known we seek a canonical realization for this
group. It turns out that the canonical realization of the
algebra of global symmetries gives, in general, a central
extension of its algebra\cite{Brown-Henneaux1}. The possibility
of a central charge cannot be ruled out by a general principle.
An example of this ``classical anomaly'' is asymptotically
anti-de Sitter 2+1 gravity\cite{Brown-Henneaux2}.  The group of
global charges in that case is the seudo-conformal group and its
canonical realization has a central charge proportional to the
inverse of the square root of the cosmological constant. We
shall see in section \ref{Global-charges} that this ``anomaly'' is
also present in Chern-Simons theory for a general group. Both,
the affine and Virasoro algebras contain non-zero central terms
providing new examples of the results reported in
\cite{Brown-Henneaux1,Brown-Henneaux2}.

The particular case of 2+1 gravity with a negative cosmological
constant is considered in section \ref{2+1-BH}.  We give
in section \ref{2+1-BH} a simple derivation of the results
reported in\cite{Brown-Henneaux2} using the Chern-Simons
formulation of 2+1 gravity.  In simple words, since the Lie
algebra $so(2,2)$ is a direct product of 2 copies of $so(2,1)$,
there are 2 commuting affine currents, one for each $so(2,1)$
copy.  Thus, by the (classical) Sugawara construction, one finds
2 copies of the Virasoro algebra, i.e., the conformal group.
The main point of section \ref{2+1-BH} is the discussion of the
necessary boundary conditions to actually obtain the 2 commuting
Virasoro algebras and to clarify the origin of the classical
central charge in them.

It is worth mentioning here that global charges have been shown
to play an important role in the definition of black hole
entropy. It is already generally accepted that the black hole
entropy is given by a boundary term in the action (at the
horizon) which is added in order to make the action
differentiable\cite{York}, i.e., it can be understood as a global charge
lying at the horizon.  An even more interesting
connection between global charges and entropy has been recently
proposed by Carlip \cite{Carlip94} in the case of 2+1
dimensions. In simple words, he computed the number of states
associated to a WZW theory defined at the black hole horizon
proving that the logaritm of this number, at least in the limit
$k\rightarrow \infty$, gives the correct value for the 2+1 black
hole entropy. The key point in Carlip's analysis is the
assumption that the horizon has to be treated as a boundary when
computing path integrals.  Whether this assumption is correct
is not yet clear, however, Carlip's result shows that those
degrees of freedom associated with the presence of boundaries
might be of central importance in the understanding of black
hole physics.

Apart from the computational simplicity of using the Chern-Simons
formulation for 2+1 gravity, the main advantage of that formalism is
the possibility to quantize the resulting algebras of global charges.
This is not an easy task when one works in the ADM formulation.
We mention some aspects of the quantization of
global charges throughout the paper.

\section{Global charges in Chern-Simons theories}
\label{Global-charges}

\subsection{Chern-Simons action}

We start with the Chern-Simons action defined in a manifold with
the topology $\Sigma \times \Re$ written in hamiltonian form,
\bb
I = \frac{k}{4\pi} \int_{\Re} dt\int_{\Sigma} d^2x \epsilon^{ij}
g_{ab} (\dot{A}^a_i A^b_j + A^a_0 F^b_{ij}) + B(\pp \Sigma)
\label{b3}
\ee
where $B$ is a boundary term that has to be included in order to
ensure gauge invariance of the action and depends on
boundary conditions.  The curvature is defined by $F^a_{ij} =
\pp_i A^a_j - \pp_j A^a_i + f^a_{\;bc} A^b_i A^c_j$, where
$f^a_{\;bc}$ are the structure constants for the Lie algebra
considered here.
The metric $g_{ab}$ is defined by  $g_{ab} = x Tr (J_a J_b)$
where $x$ is some real number depending on the representation. All
local  indices $(a,b,c,...)$
are raised and lowered with this metric which is assumed to be
invertible. We deal only with integrals over the
2-dimensional (spatial) manifold $\Sigma$ and its boundary, the
one dimensional manifold $\pp \Sigma$. Hence, all the integrands
are assumed to be 2-forms and 1-forms
respectively. We work in fundamental units $\hbar=1$ and $k$ is
dimensionless.

{}From (\ref{b3}) we learn that $A^a_0$ is a Lagrange multiplier
and the $A^a_i$ are the dynamical fields satisfying the basic Poisson
bracket
\bb
\{ A^a_i(x),A^b_j(y)\} = \frac{2\pi}{k} g^{ab} \epsilon_{ij} \delta(x,y)
\label{b4}
\ee
where $\epsilon_{ij}$ is defined as the inverse matrix of $\eij$
by $\eij\epsilon_{ik} =\delta^j_k $, no metric is needed.
The Poisson bracket of any two functions $G(A_i)$ and
$H(A_i)$ of the canonical variables is then given by
\bb
\{ G, H \} = \frac{2\pi}{k} \int \frac{\delta G}{\delta A^a_i}
\epsilon_{ij} g^{ab} \frac{\delta H}{\delta A^b_j}.
\label{b8}
\ee
The variation of (\ref{b3}) with respect to $A^a_0$ gives the
constraint
\bb
G_a \equiv \frac{k}{4\pi} g_{ab} \eij F^b_{ij} \approx 0,
\label{b5}
\ee
which satisfies the Poisson bracket algebra
\bb
\{G_a(x),G_b(y)\} = f^c_{\;ab} G_c (x) \delta(x,y).
\label{b6}
\ee
Hence, the $G_a$ are the generators of gauge
transformations acting on phase space.

All of this is well known. The question to be addressed now is
whether or not the smeared generator (supplemented with a boundary
term $Q$ that makes it differentiable\footnote{In the particular case
of gravity with the Hilbert action, the surface term is already
present in the action\cite{Hawking-Horowitz}.  This is because the
Hilbert action contains second derivatives in the metric and
therefore there is a natural surface term that is added to the
action that cancel them. In
Chern-Simons theory the action is of first order and there is no need
to add any surface term apart from those originated from demanding its
diferentiability.})
\bb
G(\eta^a) = \int_{\Sigma} \eta^a G_a + Q(\eta)
\label{b7}
\ee
satisfies a first class algebra.  It
turns out that the smeared generators provide a {\em central}
extension of the algebra of global charges.

Varying (\ref{b7}) with respect to the field $A_i$, assuming that
the parameter $\eta$ does not depend on $A_i$ in the interior,  we have
\bb
\delta G = \K\int_\Sigma \eij \eta_a D_i \delta A^a_j \
+ \ \delta Q(\eta)
\label{b10}
\ee
where $D_i v^a \equiv \pp_i v^a + f^a_{\;bc}A^b_i v^c$ is the
covariant derivative acting on a
Lie-algebra vector in the adjoint representation.
Integrating the first term by parts we find,
\bb
\delta G = - \K \int_{\Sigma} \eij D_i \eta_a \, \delta A^a_j
\ + \ \K \int_{\pp\Sigma} \eta_a \delta A^a_k dx^k + \delta Q.
\label{b11}
\ee
Therefore, if we demand the charge $Q$ to be given by
\bb
\delta Q =  - \K \int_{\pp\Sigma} \eta_a \, \delta A^a_i \, dx^i,
\label{b12}
\ee
the surface terms cancel out making the functional
derivative of $G$ well defined;
\bb
\frac{\delta G}{\delta A^a_i} =  \K \epsilon^{ij} D_j \eta_a.
\label{b13}
\ee

Using the formula (\ref{b13}) and the definition of Poisson
bracket (\ref{b8}), we can compute the Poisson bracket of two
generators smeared with parameters $\eta$ and $\lambda$
obtaining
\bb
\{ G(\eta),G(\lambda) \} = \K  \int_\Sigma  D \eta_a \w
D \lambda^a.
\label{i7}
\ee
Integrating by parts in the left hand side one obtains,
\bb
\{G(\eta),G(\lambda)\} = \int_\Sigma [\eta,\lambda]^a G_a
+ \K \int_{\pp \Sigma} \eta_a D \lambda^a
\label{i9}
\ee
where $[\eta,\lambda]^a = f^a_{\;\; bc}\eta^b \lambda^c$ is the
commutator in the Lie algebra.  Here we have used the identity
$D \w D v^a= f^a_{\;\; bc} F^b v^c $ and the definition of $G_a$ [Eq.
(\ref{b5})].

The volume term in the right hand side of (\ref{i9}) has the
expected expression as the commutator of the two parameters
times the constraint $G_a$.  However, in order to
reproduce the smeared generator, the volume term has to be supplemented
by a surface term that makes it differentiable.  This implies that the
surface term present in the right hand side of (\ref{i9}) has to be
equal to the charge that regularizes the volume term, up to a
possible central term\cite{Brown-Henneaux1}, i.e.,
\bb
\K \int \eta_a D \lambda^a  =  Q(\sigma) + K(\eta,\lambda)
\label{QK}
\ee
where $Q$ is defined by Eq. (\ref{b12}), and $\sigma$ is a parameter
that depends on $\eta$ and $\lambda$,
\bb
\sigma = \sigma(\eta,\lambda).
\label{sigma}
\ee
The precise form of the function $\sigma$ will tell us the
algebra of global charges. Of course, this depends crucially on
the chosen boundary conditions.
The term $K$, on the other hand, does not depend
on the variables that are varied at the boundary and therefore
is a central charge.  Replacing (\ref{QK}) in (\ref{i9}) one finds,
\bb
\{G(\eta),G(\lambda)\} =  G(\sigma) + K(\eta,\lambda).
\label{Alg}
\ee
It turns out that, after the gauge is fixed,
Eq. (\ref{Alg}) is still valid provide the contraints are replaced by
their corresponding charges, and the Poisson bracket is replaced by
the Dirac bracket \cite{Regge-Teitelboim,Benguria-Cordero-Teitelboim},
\bb
\{ Q(\lambda), Q(\eta) \}^* = Q(\sigma) + K(\lambda,\eta).
\label{AQ}
\ee
Therefore, if $K \neq 0$, the canonical realization of the algebra of
global charges gives a central extension of its algebra.

The purpose of this section is to study the relations
(\ref{b12}) and (\ref{QK}), extracting from them the form of
the function $\sigma(\eta,\lambda)$ and the value of $K$.
But first we have to fix the gauge and define the
set of fields at the boundary which are not ``pure gauge".

\subsection{Fixing the gauge.}
\label{Gauge-fixing}

In a field theory with no degrees of freedom like Chern-Simons
theory, the only relevant degrees of freedom are holonomies or global
charges.  As we are interested in the definition of global charges
we restrict ourselves to the special case in
which the spatial manifold has the topology of a disc. In this
section we briefly describe an appropiate way to fix the gauge in
order to keep only the degrees of freedom at the boundary.

For a disc, we call $r$ and $\phi$ the radial and polar coordinates
respectively.  The constraint dictates that the connection has to
be flat.  On the disc, this has the general solution $A_i = U^{-1} \pp_i U$
where $U$ is an arbitrary single valued group element. In order to fix the
gauge we need to impose an extra condition. We use a gauge
condition inspired by the WZW model,
\bb
A_{r,\phi}=0
\label{a3}
\ee
which implies that $U$ can be factorized in the form
\bb
U(r,\phi) = a(\phi)b(r),
\label{a4}
\ee
where $a$ and $b$ belong to the group. Substituing (\ref{a4}) in
the connection we have
\bqn
A_r(r) &=& b^{-1} \pp_r b, \\
A_{\phi}(r,\phi) &=& b^{-1}A(\phi) \, b
\label{a5}
\eqn
where $A(\phi) = a^{-1} \pp_{\phi} a$ is a function of $\phi$
only (and time).

The Lagrange multiplier $A_0$ is now found from the requirement that
the gauge fixing condition (\ref{a3}) has to be preserved by the
time evolution.  A simple calculation gives the condition over
$A_0$,
\bb
D_r(\pp_\phi A_0) = 0,
\label{A00}
\ee
where $D_r$ is the covariant derivative along the $r$
coordinate. Since the connection is flat, this equation has the
general solution
\bb
A_0 = b^{-1} \lambda(\phi) b
\label{A0}
\ee
where $\lambda(\phi)$ is an arbitrary function of the angular
coordinate $\phi$ only (and time). The arbitrary function
$\lambda(\phi)$ represents the residual gauge freedom. There is no way
to fix it by a canonical gauge condition because the constraint
is not algebraic but a differential function of the canonical
variables.  Note also that under the residual group, i.e., under
those transformations with a parameter of the form (\ref{A0}),
$A_r$ is left invariant. This follows directly from (\ref{A00})
and the definition of gauge transformation.

For later reference we also consider here the case in which the
Lagrange multiplier $A_0$ depends on the gauge field in the form,
\bb
A^a_0 = - \xi^i A^a_i.
\ee
In this case, condition (\ref{A0}) on the allowed transformations
gives the conditions over $\xi^i$,
\bb
\xi^r (\pp_r b) b^{-1} + \xi^\phi A(\phi) = \mbox{function of $\phi$ only.}
\ee
Since $A$ and $b$ are independent, this equation implies that
$\xi^\phi$ has to be a function of $\phi$ only.  Also, $b$ is not
an arbitrary function of $r$ since the term $\xi^r (\pp_r b) b^{-1}$ cannot
depend on $r$. This means that $(\pp_rb) b^{-1}$ depends on $r$ only by a
multiplicative function which is canceled by $\xi^r$. This dependence
on $r$ can be eliminated by an appropiate choice of the radial
coordinate at the boundary. Hereafter we assume that the radial
coordinate at the boundary is chosen so that
\bb
b=e^{\alpha r},  \mbox{~~~~($\alpha$: constant Lie algebra element)}
\ee
therefore $A_r=\alpha$, and $\xi^r$ is a function of the angular
coordinate only.

In summary, the gauge-fixed connection contains one arbitrary
Lie algebra value function of the angular coordinate (and time)
$A(\phi)$ defined in (\ref{a5}).  Moreover, this function transforms as a
connection under the residual gauge group defined by the
parameters of the form (\ref{A0}).

\subsection{Field independent gauge transformations. Affine Lie algebras}

In this section we study the simplest boundary condition for
which we can integrate the expression (\ref{b12}) for the
charge. Suppose that the parameter does not depend on the fields
that are varied at the boundary. In this case, the charge $Q$ is
trivially integrated from (\ref{b12}) obtaining
\bb
Q(\eta) = - \frac{k}{2\pi} \int \eta_a A^a  + Q_0,
\label{Q}
\ee
where $Q_0$ is a fixed arbitrary constant that we shall assume equal
to zero.  Now we go back to equation (\ref{QK}). The surface integral
can be rewritten as
\bqn
\K\int \eta_a D \lambda^a &=& \K\int \eta_a d \lambda^a + \K\int
                              \eta_a [A,\lambda]^a \\
                    &=& \K\int \eta_a d \lambda^a + Q([\eta,\lambda])
\eqn
where $Q$ is given in (\ref{Q}).  Therefore, in view of
(\ref{QK}), we find that in this case the function
$\sigma$ defined in (\ref{sigma}) is simply the commutator of the two
parameters
\bb
\sigma^a(\eta,\lambda) = f^a_{\; bc}\eta^b \lambda^c.
\ee
The central charge, on the other hand, is
\bb
K(\eta,\lambda) = \int_{\pp \Sigma} \eta_a d \lambda^a.
\ee
The Dirac bracket algebra of global charges is then,
\bb
\{ Q(\eta), Q(\lambda) \}^* = Q([\eta,\lambda]) + \K\int \eta_a
\, d \lambda^a
\label{i10}
\ee
where $Q$ is the gauge fixed expression for the charge,
\bb
Q= -\K \int \lambda(\phi)_a A^a(\phi)  d\phi.
\ee
Here, $A^a(\phi)$ and $\lambda(\phi)$ are, respectively, the
residual part of the connection and gauge parameters not fixed
by the gauge fixing procedure.  Thus, as expected, the canonical
realization of the residual gauge group (global symmetries)
gives a central extension of its algebra.  It is interesting to
note that the above analysis gives a geometrical meaning for the
central term in (\ref{i10}). The charges $Q$ generate the
residual gauge transformations \cite{Regge-Teitelboim} acting on
$A^a$.  If one computes the Poisson bracket between $A^a$ and
the gauge fixed generator $Q$, one finds that $A^a$ transforms
as a connection, as it should, only if the central term in
(\ref{i10}) is present.

The algebra (\ref{i10}) can be put in a more familiar form in terms
of Fourier modes. The field $A^a(\phi)$ at the boundary can be
decomposed as
\bb
A^a(\phi) = \frac{1}{k}\sum_{n=-\infty}^{\infty} T^a_n e^{in\phi}
\label{i11}
\ee
where the coefficients $T^a_n$ satisfy the ``classical'' affine algebra
\bb
\{ T^a_n,T^b_m\}^*  =  -f^{ab}_{\;\;\,c} T^c_{n+m} +  ik n g^{ab}  \,
\delta_{n+m}.
\label{i12}
\ee
Therefore, for gauge transformations whose parameters do not
depend on gauge field, the associated algebra of global charges
is the affine extension of the Lie algebra considered.

The quantum version of this algebra is obtained by
replacing the Poisson bracket by $(-i)$ times the commutator,
\bb
[T^a_n,T^b_m]  =  if^{ab}_{\;\;\;c} T^c_{n+m} + k n \, g^{ab}
\delta_{n+m}
\label{i13}
\ee
where now, the $T^a_n$ are linear operators acting on a linear
vector space. This algebra has received much study in the
context of Conformal Field Theory. The
central charge $k$ must be an integer in order to achieve gauge
invariance of the path integral under large gauge
transformations. It is remarkable that this quantization
condition also ensures the existence of highest-weight unitary
representations for the $T^a_n$.

In the context of this paper, the algebra (\ref{i12}) [or its
quantum version (\ref{i13})] has two different interpretations.
First, it arises naturally as the algebra of an infinite set of
conserved charges associated with a residual gauge symmetry at the
boundary.  A second less obvious but important interpretation is
as the basic Dirac bracket (commutator) between the basic
variables. Indeed, the basic variables after the gauge is fixed
are the components of the residual connection $A(\phi)$ (or their
Fourier components)  whose
Dirac bracket is given by (\ref{i12}). As we shall see in the next
section, for a different choice of boundary conditions, the
algebra (\ref{i12}) is no longer the algebra of conserved
charges, but still the basic Dirac bracket for the
gauge-fixed residual canonical variables.

\subsection{Field dependent gauge transformations. Virasoro algebra.}

Besides gauge transformations, Chern-Simons theory is also
invariant under diffeomorphisms. One may wonder if there exists
different boundary conditions for which the charge $Q$ represents
the freedom of making a diffeomorphism at the boundary. This is
indeed possible and it is the goal of this section.

It is well known that, in Chern-Simons theory, diffeomorphisms are
contained in the gauge group when they act on the space of solutions of
the equations of motion. Therefore, one would not expect to have an
independent set of conserved charges associated with that symmetry.
Indeed, we shall see that the charges emerging from this new class of
boundary conditions are completely determined by the ones found in
the previous section.  However, at the quantum level, normal ordering
problems appear and the algebra of global charges associated with
diffeomorphisms contain extra contributions to the central charge
that makes worth its study.

The parameters of those gauge transformations that produce
diffeomorphisms (up to the equation of motion) are of the
form\cite{Witten88}
\bb
\eta^a = -\xi^i A^a_i
\label{diff}
\ee
where $\xi^i$ is an arbitrary vector.  From this expression we
see that the parameter depends explicitly on the gauge field and
therefore, in this case, we can not straightforwardly integrate
the expression (\ref{b12}) for the charge. Since we are
interested in global charges we assume that the parameter of the
gauge transformation has the form (\ref{diff}) only at the
boundary.  As we are going to fix the gauge what happens in the
interior is immaterial.

In this section we denote the charge (associated to
diffeomorphisms at the boundary) by $J$. The variation of $J$ is
given by [see Eq. (\ref{b12})]
\bb
\delta J = \K \int \xi^i A_i \, \delta A_j dx^j .
\label{d7}
\ee
Denoting by $r$ and $\phi$ the coordinates normal and tangential to
the boundary respectively we obtain,
\bb
\delta J = \K \int \left[ \xi^r A_r \delta A_{\phi} + \m
\xi^{\phi} \delta A^2_{\phi} \right] d\phi.
\label{cd11}
\ee
In order to integrate this expression we need the boundary condition
\bb
\delta A^a_r = 0  \mbox{~~~~(at the boundary)}.
\label{d11-4}
\ee
This condition implies that $A_r^a$ at the
boundary is fixed. This is quite reasonable. Indeed, we already
know that $A_r$ at the boundary is a gauge invariant quantity
(see Section \ref{Gauge-fixing}).  We use the radial coordinate
on which $A_r^a$ is a constant Lie algebra-value element,
\bb
A^a_r \equiv \alpha^a, \mbox{~~~~~~~~$\alpha^a:$ constant Lie
algebra-value element.}
\ee

In view of (\ref{d11-4}) we obtain for the charge $J$
\bb
J(\xi^i) = \K \int \left[\xi^r \alpha A_{\phi} + \m \xi^{\phi}
A^2_{\phi} \right] d\phi + J_0 ,
\label{cd12}
\ee
where $J_0$ is an arbitrary constant whose variation vanishes
and will be adjusted later.  The formula (\ref{cd12}) gives the
value of $J$ for a single transformation with parameter $\xi^i$.
Now, we have to go back to Eq.  (\ref{QK}) and check that the
boundary term,
\bb
\K \int_{\pp \Sigma} (\xi^i A^a_i) D (\zeta^j A^b_j) g_{ab}
\label{JK}
\ee
is equal to the charge $J$ defined in (\ref{cd12}) evaluated
on some parameter $\sigma^i(\xi,\zeta)$, plus a possible central
charge.  The form of the function $\sigma$ in terms of $\xi$ and
$\zeta$ define the algebra of global charges associated to these
boundary conditions.

To prove this we first note that the covariant derivative in the
boundary term in (\ref{JK}) can be replaced by an ordinary
derivative because $Tr([A_i,A_j] A_k)=0$ for $i,j,k=1,2$.  By a simple
calculation one can prove the identity
\bb
\int (\xi^iA_i) \pp_\phi (\zeta^jA_j) d\phi =  \int \left( [\xi,\zeta]^r
\alpha A + \m [\xi,\zeta]^{\phi} A^2
\right) d\phi + \alpha^2 \int \xi^r \,\pp_{\phi} \zeta^r  d\phi
\label{cd13}
\ee
where we have already expressed $A_r$ and $A_\phi$ in terms of their residual
parts (see section \ref{Gauge-fixing}). The bracket $[,]$ denotes Lie
bracket.

{}From (\ref{cd13}) we see that the boundary term indeed contains
the charge $J$ evaluated on a parameter equal to the Lie bracket
of the deformation vectors $\xi^i$ and $\zeta^i$. This is not at
all surprising because we already knew that gauge
transformations with parameters of the form (\ref{diff})
generate diffeomorphisms.  What is more interesting is the
second term (proportional to $\alpha^2$) which, we stress, does
not depend on the variable $A(\phi)$ and therefore is a central
term in the algebra of global charges.

The algebra of global charges associated to these boundary
conditions is then
\bb
\{ J(\xi^i), J(\zeta^j) \}^* = J([\xi,\zeta]^i) +
\frac{k\alpha^2}{2\pi} \int \xi^r \,\pp_{\phi} \zeta^r  d\phi
\label{cd15}
\ee
showing that, as stated above, the
canonical realization of the symmetries at the boundary gives a
central extension of its algebra.

The Virasoro algebra is contained in (\ref{cd15}) in the special
case in which the deformation vector appearing in (\ref{diff})
has the particular form,
\bb
\xi^i =  (-\pp_{\phi} \xi , \xi),
\label{d15-5}
\ee
where $\xi=\xi(\phi)$ is an arbitrary function of $\phi$.  It is
straightforward to check that the vectors (\ref{d15-5}) form a
subalgebra in the Lie bracket. In this
case the charge (\ref{cd12}) depends only on one independent
function $\xi(\phi)$,
\bb
J(\xi) = \frac{k}{4\pi} \int \xi(\phi) \left[ \alpha^2 + 2 \alpha
A,_{\phi} + A^2 \right]
\label{Vir}
\ee
where we have already adjusted the constant $J_0$ by
\bb
J_0 = \frac{k}{4\pi} \int \xi(\phi) d\phi.
\ee

Since $A$ is a periodic function of $\phi$ we can make the Fourier
expansion
\bb
\alpha^2 + 2 \alpha A,_{\phi} + A^2 = \frac{1}{k}
\sum_{-\infty}^{\infty} L_n e^{in\phi}
\label{Fou}
\ee
and it is easy to check that the algebra (\ref{cd15}) gives the classical
Virasoro algebra for the $L_n$,
\bb
\{ L_n,L_m \}^* = i(n-m)L_{n+m} + \frac{ic_{class}}{12} n(n^2-1)\delta_{n+m}
\ee
where the central charge is given by,
\bb
c_{class} = 12 k \alpha^2.
\ee
We stress here that this central charge has a pure geometrical
origin having nothing to do with quantum normal ordering ambiguities.
In terms of Fourier modes the charge has the simple expression
\bb
J(\xi) = \sum_n L_n \xi^n
\ee
where the modes $\xi^n$ are given by
\bb
\xi^n = \frac{1}{4\pi} \int d\phi \, \xi(\phi) e^{in\phi}.
\ee
A classical central charge in the canonical realization of
global symmetries  was first found in
\cite{Brown-Henneaux2} in the context of 2+1 gravity with a negative
cosmological constant. In
the derivation given in \cite{Brown-Henneaux2} the ADM formalism was
used, and therefore, the asymptotic symmetries were defined as the
most general set of Killing vectors that leave the boundary
conditions invariant. The notion of a Killing vector requires a
metric structure which is absent in our analysis. The ad-hoc form
of deformation vectors (\ref{d15-5}) used here can be derived from
the Killing equations when a metric structure is present.

The derivation given here is general in the sense that no
particular group has been assumed. The presence of the classical
central charge is a feature that has to do only with the chosen
boundary conditions.

It should be evident that the expression for the Virasoro generators
given by (\ref{Fou}) is nothing but the (modified) Sugawara
construction. Indeed, by inverting relation (\ref{Fou}) and expressing
$A(\phi)$ also in the Fourier modes [see Eq. (\ref{i11})] we obtain,
\bb
L_n = \frac{1}{2k} \sum_m g_{ab} \, T^a_m T^b_{n-m} + in \alpha_a
T^a_n + \m k \alpha^2\delta_n
\label{Sugawara}
\ee
which can be recognized as the (modified) Sugawara construction. The
above relation is not at all surprising. The Virasoro generators
generate diffeomorphisms at the boundary while the affine
generators $T^a_n$ generate gauge transformations.  Relation
(\ref{Sugawara}) reflects the fact that diffeomorphisms in
Chern-Simons theory can be expressed as gauge transformations.  The
role of the Sugawara construction in Chern-Simons theory was first
found in \cite{Balachandran}.

The quantization of the Virasoro algebra has been extensively studied
in the literature. We refer the reader to Ref. \cite{Goddard-Olive}
for details.  We define the quantum Virasoro operator
\bb
\hat L_n = \beta :L_n: + \, a \delta_n
\ee
where $::$ means normal order.  The constant $\beta$ and $a$ are
given by
\bb
\beta = \frac{2k}{2k+Q}, \;\;\;\;\;\;\;\;\; a= \m k \alpha^2 \beta(\beta-1),
\label{beta-a}
\ee
and $Q g^{ad}=f^a_{\;\;bc}f^{dbc}$ is the quadratic Casimir in the adjoint
representation.  The operator $\hat L_n$ satisfies the quantum
Virasoro algebra
\bb
{}~[ \hat L_n, \hat L_m ] = (n-m) \hat L_{n+m} + \frac{c}{12}
n(n^2-1)\delta_{n+m}
\ee
with a total central charge,
\bb
c =  12 k \alpha^2\beta^2 + \beta N .
\label{cT}
\ee
Here $N$ is the dimension of the Lie algebra considered. The second
term in this expression is the well known central charge induced by
the Sugawara construction and has a pure quantum origin.  The first
term, on the other hand, has to do only with the chosen boundary
conditions. Note that the classical contribution appears in
(\ref{cT}) multiplied by $\beta^2$.


\section{SO(2,2) gravity and the 2+1 black hole}
\label{2+1-BH}

\subsection{Chern-Simons formulation for 2+1 Gravity}

Three dimensional gravity (with a negative cosmological
constant) can be written as a Chern-Simons theory for the
anti-de Sitter group\cite{Achucarro,Witten88}. The goal of this
section is to describe how the results of the previous section
can be applied to the case of 2+1 gravity. We shall be
particularly interested in the quantization of the 2+1 black hole
mass and angular momentum defined as global charges associated
with symmetries of the boundary\footnote{See
\cite{Cangemi} for an equivalent definition for these charges
in terms of holonomies. Global charges in Chern-Simons gravity
without a cosmological constant has been studied in \cite{Bimonti}.}.

The gauge field for 2+1 gravity with a negative cosmological constant is
\bb
A= e^a P_a + w^a J_a
\ee
where $P_a$ and $J_a$ satisfy the $so(2,2)$ algebra
\bqn
{}~[J_a,J_b] &=& \epsilon_{ab}^{\ \ c} J_c \nonumber\\
{}~[J_a,P_b] &=& \epsilon_{ab}^{\ \ c} P_c \\
{}~[P_a,P_b] &=& \frac{1}{l^2}\epsilon_{ab}^{\ \ c} J_c \nonumber .
\eqn
The 1-forms $e^a, w^a$ are the triad and spin connection
respectively. The generators $J_a^{\pm} = \m (J_a \pm lP_a) $
satisfy both the algebra of $so(2,1)$ and commute between them.
The corresponding connections for each $so(2,1)$ copy are,
\bb
A^a_\pm = w^a \pm \frac{e^a}{l}.
\ee
In terms of these new fields the Chern-Simons action splits in
the sum of two Chern-Simons actions
\bb
I(e,w) = I^+(A_+) + I^-(A_-),
\label{cs2}
\ee
where
\bb
I^\pm(A^a) = \pm \frac{k}{4\pi} \int (\eta_{ab} A^a \w d A^b +
(1/3) \epsilon_{abc} A^a \w A^b \w A^c )
\ee
with $\eta_{ab} = diag(-1,1,1)$ and $\epsilon_{012}=1$.  We choose
the coupling constant $k$ equal to
\bb
k = \frac{l}{G}
\ee
in order to agree with the conventions followed in \cite{BTZ}. In
fundamental units $\hbar=G=c=1$, $k$ is dimensionless.

All the results described in previous sections can be applied
to the present case which consists of two copies of the
Chern-Simons action.
It should be kept in mind that the sector $(-)$ has a negative
coupling constant $(-k)$.  The Hamiltonian is given by
\bb
H(A_0) = \int (A^+_0)^a \, G_a(A_+) + B^+ - \int (A^-_0)^a \, G_a(A_-) - B^-
\ee
where $B^\pm$ are boundary terms that regularize each sector
independently.

\subsection{The 2+1 black hole}

The Einstein equations in 2+1 dimensions with a negative
cosmological term possess a black hole solution\cite{BTZ}. This solution,
in the absence of electromagnetic fields, is
characterized by its mass and angular momentum. The metric takes
the form,
\bb
ds^2 = -N^2 dt^2 + N^{-2} dr^2  + r^2(N^{\phi} dt + d\phi)^2,
\label{bh1}
\ee
where the squared lapse $N^2(r)$ and the angular shift
$N^{\phi}(r)$ are given by
\begin{eqnarray}
N^2(r) &=& - M + \frac{r^2}{l^2} + \frac{J^2}{4r^2} \label{2.2} \nonumber \\
N^{\phi}(r) &=& - \frac{J}{2r^2} \nonumber
\end{eqnarray}
with $-\infty < t <\infty$,  $0 < r < \infty$  and  $0 \leq \phi \leq 2\pi$.

The lapse function $N(r)$ vanishes for two values of $r$ given by
$$
r_{\pm}=l \left[ \frac{M}{2} \left( 1 \pm \sqrt{1-\left(
\frac{J}{Ml}\right)^2} \right) \right]^{1/2}.
$$
Of these, $r_{+}$ is the black hole horizon.  The relation
between $r_\pm$ and $M,J$ can be inverted obtaining the useful
formulae,
\bb
M = \frac{r_+^2 + r_-^2}{l^2}, \;\;\;\;\;\; J= \frac{-2r_+ r_-}{l}.
\label{MJ}
\ee
In order for the horizon to exist one must have
\bb
M>0,  \;\;\;\;  |J| \leq Ml.
\label{CCS}
\ee
In the extreme case $|J|=Ml$, both roots of $N^2=0$ coincide.

The metric (\ref{bh1}) has two integration constants $r_+$ and
$r_-$, or, equivalently, $M$ and $J$.
There are 2 additional parameters associated with the freedom of
multiplying the lapse function $N$ by an arbitrary constant
$N_\infty$, and the freedom to add to $N^\phi$ and arbitrary
constant $N^\phi_\infty$. These freedoms are global
symmetries whose conserved charges are precisely the mass $M$ and the
angular momentum $J$ respectively. In the ADM Hamiltonian
variational principle \cite{BHTZ} one keeps $N_\infty$ and
$N^\phi_\infty$ fixed at the boundary while $M$ and $J$ are
varied.  The purpose of this section is to translate the ADM
variational principle to the Chern-Simons language. As we shall
see, the Chern-Simons formulation allows for great
simplifications.

For our purposes here it is useful to introduce a Rindler-like
radial coordinate for the black hole metric. For $r \geq r_+$ we
define a dimensionless radial coordinate $\rho$ by
\bb
r^2 = r_+^2 \cosh^2 \rho - r_-^2 \sinh^2 \rho   \;\;\;\;\;\; (\rho>0).
\label{rho}
\ee
This radial coordinate brings the metric into the
simple form
\bb
ds^2 = -\sinh^2\rho [r_+ d\tau - r_- d\phi]^2 + l^2 d\rho^2 +
\cosh^2\rho [-r_- d\tau + r_+ d\phi]^2
\label{ads}
\ee
where we have defined the dimensionless time coordinate $\tau=t/l$.
[The coordinate $\rho$ can be extended to the black hole
interior by replacing the hyperbolic functions by appropiated circular
functions. Since we are interested in the metric at the outer
boundary, we do not need to do this here.] It should be evident
from (\ref{ads}) that the black hole differs from anti-de Sitter
space only in its global properties\cite{BHTZ}.

{}From (\ref{ads}) we can read the form of the triad (up to local
Lorentz rotations)
\bqn
e^0 &=& (r_+ d\tau - r_- d\phi) \sinh\rho \nonumber\\
e^2 &=& (-r_- d\tau + r_+ d\phi) \cosh\rho \label{e}\\
e^1 &=& ld\rho, \nonumber
\eqn
and, solving the torsion equation $T^a=0$, one finds the spin
connection (up to a Lorentz rotation),
\bqn
w^2 &=&(1/l) (r_+ d\tau - r_- d\phi) \cosh\rho \nonumber\\
w^0 &=&(1/l) (-r_- d\tau + r_+ d\phi) \sinh\rho  \label{w}\\
w^1 &=& 0. \nonumber
\eqn
This solution corresponds to the ``zero mode'' solution
considered before. Indeed, as discussed in section
\ref{Gauge-fixing}, the general form of the solution for
the constraints is $A_\phi= b^{-1} A(\phi) b$ and $A_r =
\alpha$, where $b=e^{\alpha \rho}$ is a group element that only depends on the
radial component. The above black hole solution corresponds to the
simple case when the Lie algebra value function $A(\phi)$
is constant. The right $(A^+)$ and left
$(A^-)$ functions associated to the black hole can be easily
calculated.  They  are,
\bqn
A^+ &=& \frac{r_+ - r_-}{l} J_2^+, \label{A+}\\
A^- &=& - \frac{r_+ + r_-}{l} J_2^- \label{A-}
\eqn
On the other hand, he group elements $b_\pm$ are given by
\bb
b_\pm = e^{\pm \rho J^\pm_1}
\ee
therefore $ A^\pm_r \equiv \alpha_\pm = \pm J^\pm_1$.

\subsection{The black hole charges. Mass and angular momentum.}

In this section we shall prove that the mass $(M)$ and the angular
momentum $(J)$ of the black hole solution are contained in the
charges defined in Sec. \ref{Global-charges} when one
goes to the $SO(2,2)$ group.

The black hole charges make use of the fact that the Lie algebra
$so(2,2)$ splits into 2 copies of $so(2,1)$.
Indeed, the Chern-Simons variational principle that gives rise
to the ``zero mode'' black hole solution allows for
independent diffeomorphisms in each copy of the Chern-Simons
action (\ref{cs2}).

In other words, we can consider the boundary condition for the
Lagrange multiplier $A_0 = A_0^+ + A_0^-$
\bb
A_0^\pm = \mp \xi_\pm^i A^\pm_i  \;\;\;\;\;\;\;\ \mbox{(at the boundary).}
\label{B2}
\ee
where $\xi_+^i $ and $\xi_-^i $ are 2 independent vectors of the
form (\ref{d15-5}).  It should be stressed here that these boundary
values for $A_0$ double the number of parameters in the global
gauge group.

These boundary values for $A_0$ give us 2
copies of the Virasoro algebra given by [see Eq. (\ref{Vir})]
\bb
L^\pm_n = \frac{k}{4\pi} \int e^{-in\phi} \left( \alpha^2_\pm +
2\alpha_\pm \pp_\phi A_\pm + A_\pm^2 \right),
\label{Vir-Grav}
\ee
which, both, satisfy the classical Virasoro algebra with a
central charge equal to
\bb
c_\pm = 12 k\alpha^2_\pm.
\ee

The values of $A_\pm$ for the black hole are
summarized in (\ref{A+}) and (\ref{A-}). Since for the
black hole $A_\pm$ does not depend on $\phi$ the only non-zero
charge is the zero mode. By direct replacement we find
\bb
L_0^\pm = \frac{k}{2} ( 1 + M \pm J/l )
\ee
where $M$ and $J$ in terms of $r_\pm$ are given in (\ref{MJ}).
Therefore, we have
\bqn
M + 1 &=& (L^+_0 + L^-_0)/l \label{M0} \\
J &=& L^+_0 - L^-_0 .
\label{M}
\eqn
Hence, we see that indeed the black holes charges $M$ and $J$
are related to the zero modes charges studied in Sec.
\ref{Global-charges}.  Similarly, the zero-mode component of
the functions $\xi_\pm(\phi)$ are related to the ADM ``lapse''
$(N_\infty)$ and ``shift'' $(N^\phi_\infty)$ functions by,
\bb
\xi_\pm^0 = N_\infty \pm lN^\phi_\infty.
\ee
We remember here that in the ADM Hamiltonian variational principle
$N_\infty$ and $N^\phi_\infty$ are held fixed
while the mass and angular momentum are
varied\cite{BHTZ}. In the present Chern-Simons variational
principle we have kept fixed the deformation vectors $\xi_\pm$ while
the function $A(\phi)$ is varied. Thus, both variational
principles are completely equivalent.

It should be noted that our expression for $M$ in terms of
the zero-mode Virasoro generators differs from the one exhibited in
\cite{BHTZ} by the additive constant $+1$. The convention followed in
this paper was to adjust the zero point of the energy so that the
charges satisfy the Virasoro algebra with the standard expression for
the central charge, i.e., having and $SL(2,\Re)$ sub-algebra in its
centre.

The conformal group of asymmptotic motions for anti-de Sitter 2+1
gravity (two copies of the Virasoro algebra) and the presence of a
classical central charge was first found in \cite{Brown-Henneaux2}
using the ADM formalism.  The simplicity of our derivation based in
the Chern-Simons formulation for 2+1 gravity should be noted.

\subsection{Quantization}

Lastly we consider the quantization of the above algebra. As discussed
before, the classical Virasoro algebra cannot be quantized
straightforwardly because of normal ordering problems.  The Virasoro
generators (\ref{Vir-Grav}) have to be modified by
\bb
\hat L^\pm_n = \beta_\pm : L^\pm_n : + \, a_\pm \delta_n
\label{hat-L}
\ee
where both generators satisfy the Virasoro algebra with central
charge equal to
\bb
c_\pm = \beta_\pm N + 12 k \beta^2_\pm \alpha_\pm^2 .
\ee
In our normalizations the quadratic Casimir is given by $Q=-2$.
Therefore, the constants $\beta_\pm$ and $a_\pm$ are given by [see
Eq. (\ref{beta-a})]
\bb
\beta_\pm = \frac{k}{k \mp 1}, \;\;\;\;\;\;\;\;\;\; a_\pm =
\frac{k^2}{2(k \mp 1)^2}.
\ee
The relation between the zero-mode operators $L_0^\pm$ with the mass
and angular momentum  of the black hole [Eqs. (\ref{M0}) and
(\ref{M})] suggests that only
highest-weight representations for the Virasoro generators should be
physically relevant. These representations have a lower bound state
$|\omega>$ satisfying
\bb
\hat L_n |\omega> = 0,   \;\;\;\;\;  T^a_n|w>=0  \;\;\;\;\;\;\;    (n>0).
\ee
The value of the operators $\hat L^\pm_0$ on this state are
\bb
\hat L^\pm_0 = \frac{k^2}{2(k\mp 1)^2}[k + (k \mp 1)(M \mp J/l)].
\ee
Due to the shift $a_\pm$ in (\ref{hat-L}), a reasonable
condition for the lowest eigenvalue of $\hat L^\pm_0$ is
\bb
\hat L^\pm_0 - a_\pm \geq 0.   \\
\label{>0}
\ee
This condition gives the following bounds for the black-hole mass
and angular momentum,
\bb
M + 1 \geq |J/l|.
\label{bound}
\ee
The black hole horizon exists when $M>|J/l|$, therefore
condition (\ref{bound}) is not enough to ensure cosmic
censorship.  In the context of this paper this is irrelevant
because one can always add a constant to $L_0$ shifting its zero
point.  This shift, however, alters the form of the central
charge.  The criterium followed in this paper has been to keep
the usual form of the central charge in the Virasoro algebra
obtaining the bound (\ref{bound}). As discussed in \cite{BTZ}, the
region $M>0$ correspond to the black hole spectrum (with a
causal singularity \cite{BHTZ} at the origin), and $M=0$ is the
black hole vacuum. On the other hand, the region $-1<M<0$
corresponds to the point particles (with conical singularities)
discussed in\cite{Deser-Jackiw}.  Finally the lower state,
$M=-1$, is anti-de Sitter space which has no singularities.

Similars bounds for the possible values of $M$ and $J$ has been
discussed in \cite{Coussaert-Henneaux,Townsend} using
supersymmetric techniques.  A proof of an energy theorem in 2+1
gravity is given in \cite{Menotti}.

\noindent {\bf Acknowledgements}

I would like to thank S. Carlip, L.J. Garay, M.  Henneaux,
A.  Mikovi\'{c}, C. Teitelboim, and J. Zanelli for many enlightening
conversations; K.S. Gupta and G. Bimonti for bringing their work to my
attention,
and M.B. Halpern for useful correspondence. This work
was partially supported by grants 1930910-93 and 1940203-94 from
FONDECYT (Chile), a grant from Fundaci\'on Andes (Chile), and by institutional
support to the Centro de Estudios Cient\'{\i}ficos de Santiago
provided by SAREC (Sweden) and a group of chilean private companies
(COPEC,CMPC,ENERSIS).


\begin{references}

\bibitem{CS-WZW} E. Witten, {\em Commun. Math. Phys.} {\bf
121}, 351 (1989),  G. Moore and N. Sieberg, {\em Phys. Lett.}
{\bf B220}, 422 (1989); S. Elizur, G. Moore, A. Schwimmer and
N. Sieberg {\em Nucl. Phys.} {\bf B326}, 108 (1989).

\bibitem{Witten84} E. Witten, {\em Commun. Math. Phys.} {\bf 92}, 455 (1984).

\bibitem{Balachandran} A. P. Balachandran, G. Bimonti, K.S.
Gupta, A. Stern, \IJMPA{7}, 4655 (1992).

\bibitem{Ezawa} K. Ezawa, OU-HET/206, November 1994, preprint.

\bibitem{Regge-Teitelboim} T. Regge and C. Teitelboim, {\em Ann.
Phys.} (N.Y.) {\bf 88}, 286 (1974).

\bibitem{Brown-Henneaux1} J.D. Brown and M. Henneaux, {\em Journ.
Math. Phys.} {\bf 27}, 489 (1986).

\bibitem{Brown-Henneaux2} J.D. Brown and M. Henneaux, {\em Commun.
Math. Phys.} {\bf 104}, 207 (1986).

\bibitem{York} J.D. Brown, E.A. Martinez and J.W.York,
\PRL{66}, 2281 (1991); R.M. Wald, \PRD{48}, R3427, (1993); M.
Ba\~nados, C. Teitelboim and J. Zanelli, \PRL{72}, 957 (1994).

\bibitem{Carlip94} S. Carlip, ``The statistical mechanics of the 2+1
black hole" UCD-94-32, NI-94011, preprint.

\bibitem{Hawking-Horowitz} S. W. Hawking and G. T. Horowitz, DAMTP/R
94-52, UCSBTH-94-37, preprint.

\bibitem{Benguria-Cordero-Teitelboim} R. Benguria, P. Cordero and C.
Teitelboim, {\em Nucl. Phys.} {\bf B122}, 61 (1977).

\bibitem{Witten88} E. Witten, {\em Nucl. Phys.} {\bf B 311}, 46 (1988).

\bibitem{Goddard-Olive} For a review see, P. Goddard and D. Olive,
{\em Inter. Journ. of Mod. Phys.} {\bf A1}, 303, (1986).

\bibitem{Achucarro} A. Ach\'ucarro and P.K. Townsend, \PLB{180},
89 (1986).

\bibitem{Cangemi} D. Cangemi, M. Leblanc and R.B. Mann,
\PRD{48}, 3606 (1993).

\bibitem{Bimonti} G. Bimonti, K.S. Gupta, A. Stern, \IJMPA{8}, 653
(1993).

\bibitem{BTZ} M. Ba\~nados, C. Teitelboim and J.Zanelli,
\PRL{69}, 1849 (1992).

\bibitem{BHTZ} M. Ba\~nados, M. Henneaux, C. Teitelboim and
J.Zanelli, \PRD{48}, 1506 (1993).


\bibitem{Deser-Jackiw} S. Deser, R. Jackiw and G. 't Hooft,
\Ann{152}, 220 (1984).

\bibitem{Coussaert-Henneaux} O. Coussaert and M. Henneaux,
\PRL{72}, 183 (1994).

\bibitem{Townsend} J. M. Izquierdo and P.K. Townsend, DAMTP
preprint, December 1994.

\bibitem{Menotti} P. Menotti, MIT-CTP \# 2324, IFUP-TH-33/94,
preprint.

\end{references}
\end{document}